\documentclass[aoas,nameyear,MSNbibl,dvips]{arximspdf}
\usepackage{graphicx}

\doi{10.1214/10-AOAS398C}
\referstodoi{10.1214/10-AOAS398}
\volume{5}
\issue{1}
\pubyear{2011}
\firstpage{52}
\lastpage{55}

\begin{document}
\begin{frontmatter}

\title{Discussion of: A statistical analysis of multiple
temperature proxies: Are reconstructions of surface temperatures over
the last\\ 1000
years reliable?}
\runtitle{Discussion}
\pdftitle{Discussion on A statistical analysis of multiple temperature
proxies: Are reconstructions of surface temperatures over the last 1000
years reliable?
by B. B. McShane and A. J. Wyner}

\begin{aug}
\author{\fnms{Richard A.} \snm{Davis}\corref{}\thanksref{t1}\ead[label=e1]{rdavis@stat.columbia.edu}}
and
\author{\fnms{Jingchen} \snm{Liu}\thanksref{t2}\ead[label=e2]{jcliu@stat.columbia.edu}}

\thankstext{t1}{This research was supported in part by NSF Grant DMS-07-43459.}
\thankstext{t2}{This research was supported in part by the Institute of
Education Sciences, U.S. Department of Education Grant R305D100017.}

\runauthor{R. A. Davis and J. Liu}

\affiliation{Columbia University}

\address{Department of Statistics\\
Columbia University\\
New York, New York 10027\\
USA\\
\printead{e1}\\
\phantom{E-mail:\ }\printead*{e2}}
\end{aug}

\received{\smonth{9} \syear{2010}}



\end{frontmatter}

\section{Introduction}

It is a pleasure to have the opportunity to read and comment on McShane
and Wyner's paper, ``A statistical analysis of multiple temperature
proxies.'' This is a must read for every statistician who has an
interest in the climate change debate that continues to be a source of
intense public policy discussions. The authors are to be congratulated
for writing a clear and accessible article that helps decipher the
statistics behind the scientific claims related to the
paleoclimatological side of the issue.

\begin{figure}

\includegraphics{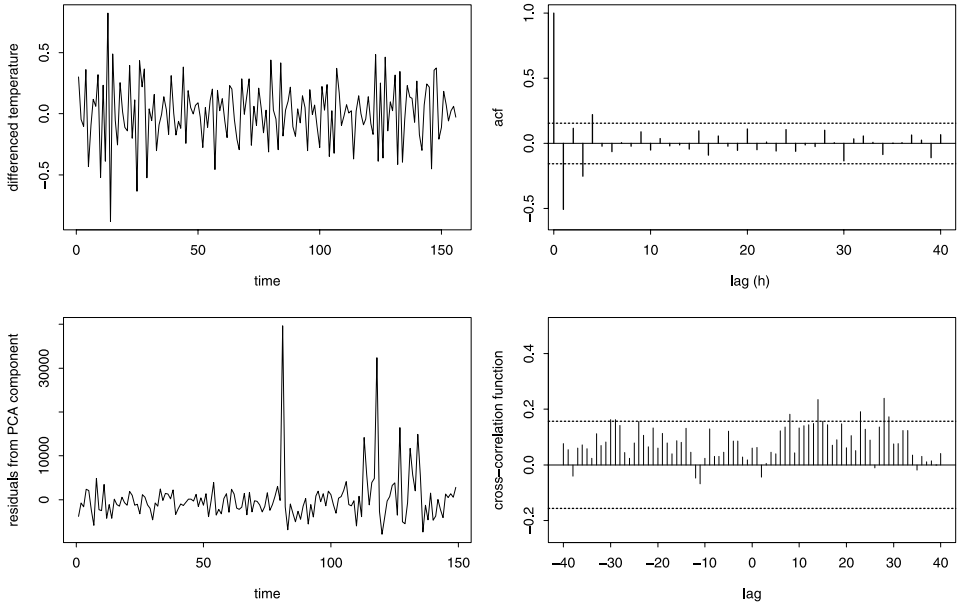}

\caption{Upper panel shows time series plot and ACF of differenced
temperature series; lower panel shows residuals from time series fit to
the first component in the PCA decomposition of the proxies and the
corresponding cross-correlation function with the temperature series.
The code for producing these graphs can be found in Davis and Liu (\protect\citeyear{davisliu}).}
\label{fig1}
\end{figure}

We will focus our discussion on some of the points dealing with the
time series modeling aspects. The main objectives presented in this
paper are strategies for selecting and evaluating predictive models of
average yearly temperature that include nearly 1200 proxies. For the
sake of this discussion, we will concentrate on the response consisting
of the CRU Nothern Hemisphere annual mean land temperature (upper-left
panel of Figure 5 in the paper) from 1850 to 1999. Roughly, one can
discern three or possibly four segments in this time series: the first
from 1850 to 1920 with nearly constant mean, the second from 1921 to
1970 with a mean that is increasing slightly, and the third from 1971
to 1999 with a sharply increasing mean. This is roughly consistent with
the three segments found by the segmentation program AutoPARM,
developed by Davis, Lee anf Rodriguez-Yam~(\citeyear{davisleeyam}). If we let $Y_1,\ldots
,Y_{150}$ denote the temperature data during these 150 years,
1850--1999, the differenced series $\nabla Y_t=Y_t-Y_{t-1}$ and its
autocorrelation function (ACF) are plotted in the upper-left and right
panels of Figure \ref{fig1}. The differenced series looks stationary and the ACF
has a spike of $-$0.5 at lag 1, has small values for lags 2 and 3 and is
essentially 0 for lags greater than 4. This ACF has the signature of a
classical moving average time series with a unit root. Such an ACF
suggests a model that takes the form
\[
Y_t=X_t+Z_t ,
\]
where $\{Z_t\}$ is IID with mean 0 and variance $\sigma^2$ and the signal $X_t$
is slowly moving with $\nabla X_t=X_t-X_{t-1}$ being small and having
little temporal dependence. If one views the signal as a proxy for the
regression function consisting of linear combinations of \textit{proxies},
then there is just not much signal present. This is consistent with
McShane and Wyner's observation that ``the temperature signal in the
proxy record is surprisingly weak.'' So in this case, the simple
diagnostic of looking at~the ACF of the differenced series has revealed
a great deal about the structure of the~time series. In particular, it
more than likely excludes a random walk model for the data.

In building forecast models, the inclusion of proxies or covariates is
not always straightforward. Often, and as illustrated in this paper, a
pure time series model can be as effective for forecasting future (or
past) values as one that includes a range of covariates. In fact, a
particular covariate that is independent of the response, but is able
to mimic the dependence structure of the response, can lead to spurious
results. Given the nature of the temperature time series, and the
seemingly poor performance of the \textit{optimal} linear models under
consideration, one wonders if a more sophisticated time series modeling
approach that incorporates nonlinear effects in a few well-chosen
covariates would be more effective. To illustrate this point, consider
only one covariate corresponding to the component\vadjust{\goodbreak} with largest variance
in the PCA decomposition based on the 1209 covariates in the period
1851--2000. The residuals $\hat u_t$ in fitting an ARMA model to this
data are displayed in the lower-left panel of Figure \ref{fig1}. Notice the two
large \textit{outliers} occurring at times 1930 and 1970 as well as a
possible increase in variance during the last 30 years. As rightly
pointed out in the McShane--Wyner paper, in looking for significant
cross-correlation in a time series, one of the component series should
first be whitened. In this case we whitened the PCA factor and computed
the cross-correlation between $\hat u_t$ with the temperature time
series $Y_t$. The sample correlations between $Y_{t+h}$ and $\hat u_t$,
for lags $h=-40,\ldots,40$, are displayed in the lower-right panel of
Figure \ref{fig1} together with 95\% confidence bounds based on the two series
being uncorrelated. At lag zero, there is virtually no correlation
between the two series. The largest and significant correlations occur
at lags $h=14$ and 28, suggesting a period of around 14 years in the
dependence between the two series. So in building a regression model of
the form used in Section 5.1 of the McShane--Wyner paper, it may make
more sense to lag the PCA component by $-14$ instead of using the
contemporaneous component. This brings up the entire issue of time
synchronization with the proxies, which seems to have been ignored
entirely in this paper's discussion. It is puzzling that we find a
lagged effect in which the covariate leads (happens before) the
response. Based on physical considerations, it would seem that proxies
should follow rather than lead temperature. It is easy to construct
scenarios in which a proxy provides no forecasting information when
only included contemporaneously, but has a strong predictive effect
when lagged. In any case, lagged effects with the covariates should be
more fully explored and included as potential \textit{covariates} in the model.

It would be worth exploring potential connections between outliers in
the covariate series corresponding with other features in the
temperature series. For example, the PCA residual time series graphed
in the lower-left panel of Figure \ref{fig1} displays outliers at years 1931 and
1968 that are relatively close to the times at which we noticed a
structural break in the actual temperature series. Each of these years
corresponds to a change in slope of the \textit{linear trend}. While we
are not suggesting such a simplistic model for the temperature series,
this example suggests a range of models that may carry more predictive
skill. We think the interesting comparison would be between the
regression models with covariates chosen from a large set of proxies
versus a transfer-style model [see, e.g., Brockwell and Davis (\citeyear{bd2002})]
that uses a handful of strategically selected covariates. Transfer
functions are also amenable for capturing intervention effects and can
serve as a starting point for capturing nonlinear effects between the
response and proxies.

Certainly, the paleoclimatological reconstruction problem offers a
difficult modeling challenge to climatologists and statisticians alike.
It will be fascinating to see the fruits of these efforts in years to come.

%


\begin{supplement}
\stitle{R-code for computing the graphs displayed in Figure~\ref{fig1}\\}
\slink[doi]{10.1214/10-AOAS398CSUPP}
\slink[url]{http://lib.stat.cmu.edu/aoas/398C/supplementC.txt}
\sdatatype{.txt}
\sdescription{This code can be used to reproduce the graphs displayed
in Figure \ref{fig1}.}
\end{supplement}


%

\printaddresses

\end{document}